\documentstyle[aps,epsf,epsfig]{revtex}
\newcommand{\bq}{\begin{equation}}
\newcommand{\eq}{\end{equation}}
\newcommand{\bqn}{\begin{eqnarray}}
\newcommand{\eqn}{\end{eqnarray}}
\newcommand{\nb}{\nonumber}
\newcommand{\lb}{\label}

\begin{document}
\title{Critical Collapse of  Cylindrically Symmetric Scalar Field
 in Four-Dimensional Einstein's Theory of Gravity}
\author{ Anzhong Wang  \thanks{E-mail: Anzhong$\_$Wang@baylor.edu.
On leave from Department of Theoretical Physics, the State
University of Rio de Janeiro, Brazil.}}
\address{CASPER, Physics Department, P.O. Box 97316, Baylor University,
Waco, TX76798-7316\\
Department of Physics, University of Illinois at Urbana-Champaign,
1110 West Green Street, Urbana, IL61801-3080}

\date{\today }
\maketitle

\begin{abstract}

Four-dimensional cylindrically symmetric spacetimes with homothetic
self-similarity are studied in the context of Einstein's Theory of
Gravity, and a class of exact solutions to the Einstein-massless scalar
field equations is found. Their local and global properties are
investigated and found that they represent gravitational collapse
of a massless scalar field. In some cases the collapse forms  black
holes with cylindrical symmetry, while in the other cases it does not.
The linear perturbations of these solutions are also studied and given
in closed form. From the spectra of the unstable eigen-modes, it
is found that there exists one solution that has precisely one unstable
mode, which may represent a critical solution, sitting on a boundary
that separates two different basins of attraction in the phase space.

\end{abstract}

\vspace{.6cm}


\section{Introduction}
\lb{SecI}
\renewcommand{\theequation}{1.\arabic{equation}}
\setcounter{equation}{0}

The studies of non-linearity of the Einstein field equations near
the threshold of black hole formation reveal very rich phenomena
\cite{Chop93},
which are quite similar to critical phenomena in Statistical
Mechanics and Quantum Field Theory \cite{Golden}. In particular,
by numerically studying the gravitational collapse of a massless
scalar field in $3+1$-dimensional spherically symmetric
spacetimes, Choptuik found that the mass of such formed black
holes takes the form,
\bq
\lb{1.1}
  M_{BH} = C(p)\left(p -p^{*}\right)^{\gamma},
\eq
where $C(p)$ is a constant and depends on the initial data, and
$p$ parameterizes a family of initial data in such a way that when
$p > p^{*}$  black holes are formed, and when $p < p^{*}$ no black
holes are formed. It was shown that, in contrast to $C(p)$, the
exponent $\gamma$   is {\em universal} to all the families of initial
data studied, and was numerically  determined as $\gamma \sim 0.37$.
The solution with $p = p^{*}$, usually called the critical
solution, is found also {\em universal}. Moreover, for the massless
scalar field it is periodic, too. Universality  of the critical
solution and the exponent $\gamma$, as well as the power-law scaling of the
black hole mass all have given rise to the name {\em Critical
Phenomena in Gravitational Collapse}.

Choptuik's studies were soon generalized to other matter fields
\cite{Gun00}. From all the work done so far,
the following seems clear: (a) There are two types of critical
collapse, depending on whether the black hole mass takes the
scaling form (\ref{1.1}) or not. When it takes the  form,
the corresponding collapse is called Type $II$ collapse, and when
it does not it is called Type $I$ collapse. In the type $II$
collapse, all the critical solutions found so far have either
discrete self-similarity (DSS) or homothetic self-similarity
(HSS), depending on the matter fields. In the type $I$ collapse,
the critical solutions have neither DSS nor HSS. For certain
matter fields, these two types of collapse can co-exist. (b) For
Type $II$ collapse, the corresponding exponent is universal only
with respect to certain matter fields. Usually, different matter
fields have different critical solutions and
different exponents. But for a given matter field the critical
solution and the exponent are universal \footnote{So far, the
studies have been mainly restricted to spherically symmetric case
and their non-spherical linear perturbations \cite{Gun00}. Therefore, it is not
really clear whether or not the critical solution and exponent are
universal with respect to different symmetries of the
spacetimes.}. (c)  A critical solution for both of the two types
has {\em one and only one unstable mode}. This now is considered as one
of the main criteria for a solution to be critical. (d) The
universality of the exponent is closely  related to the number of unstable
modes. In fact, the unstable mode, say, $k_{1}$, of the critical solution
is related to the exponent $\gamma$ via the relation,
$\gamma =  \left|k_{1}\right|^{-1}$,
which can be obtained by using dimensional analysis \cite{Even}.

From the above, one can see that to study
critical collapse, one may first find some particular solutions by
imposing certain symmetries, such as, DSS or HSS. This can
simplify the problem considerably. For example, in the spherically symmetric
case, by imposing HSS symmetry the Einstein field equations will be
reduced from PDE's to ODE's.    Once the particular solutions are
known, one can study their linear perturbations and find out the
spectrum of the corresponding eigen-modes. If a solution has  precisely
one unstable mode,  it may represent a  critical solution,
sitting on a boundary that separates
two different basins of attraction in the phase space.

The studies of critical collapse  have been
mainly numerical so far, and analytical ones are still highly
hindered by the complexity of the problem, even after imposing
some symmetries. Lately, some progress has been achieved in the
studies of critical collapse of a massless scalar field in an anti-de
Sitter background in $2+1$-dimensional spacetimes
both numerically \cite{PC00,HO01} and analytically \cite{Gar01,CF01,HW02}.
This serves as the first analytical model in critical collapse.

In this paper, we shall present another analytical model that
represents critical collapse of a massless scalar field
in four-dimensional Einstein's Theory of Gravity with cylindrical symmetry.
Although spacetimes with cylindrical
symmetry do not represent realistical models, the studies of them
can provide deep insight into the nonlinearity of the Einstein field
equations. In particular, they may shine some light on the
possible roles that gravitational radiation and angular momentum may play
in critical collapse.  In fact, such studies have already shown to be
very useful in probing   non-spherical gravitational collapse \cite{AT92}.
In addition, they may also provide a useful
testbed for numerical relativity \cite{d'In} and Quantum Gravity \cite{Ast}.

The rest of the  paper is organized as follows: In Sec. II we
first review the regularity
conditions for  a four-dimensional cylindrical spacetime, including the ones
at the symmetry axis. Then we introduce the notion of homothetic self-similarity
with cylindrical symmetry. In Sec. III, a  class of exact solutions
with such a symmetry to the Einstein-massless scalar field equations is
presented. It is shown that
they represent gravitational collapse of a scalar field, in which black
holes can be formed.  In Sec. IV, the linear perturbations of these solutions are
studied and given in closed form.
After properly imposing boundary conditions, the spectra of the unstable
modes of the perturbations are determined. In particular, it is found that there
exists a solution that has precisely one unstable mode, which may represent
a critical solution, sitting on a boundary that separates
two different basins of attraction in the phase space.
In Sec. V, the main results are summarized and some concluding remarks are
given. There are also two appendices, $A$ and $B$. In Appendix $A$, the Ricci
tensor is given in terms of self-similar variables. The linear terms of perturbations
of the Ricci tensor are also given there. In Appendix $B$, the expansions of the
out- and in-going radial null geodesics are calculated, from which trapped surfaces
and apparent horizons are defined.

\section{Spacetimes with Homothetic Self-Similarity}
\lb{SecII}
\renewcommand{\theequation}{2.\arabic{equation}}
\setcounter{equation}{0}

The general metric for cylindrical spacetimes with two hypersurface orthogonal Killing
vectors takes the form \cite{Kramer80},

\bq
\lb{2.1}
ds^{2} = e^{-M(t,r)}\left(dt^{2} - dr^{2}\right)
- r^{2}e^{-S(t,r)}\left(e^{V(t,r)}dw^{2}
+ e^{-V(t,r)}d\theta^{2}\right),
\eq
where $x^{\mu} = \{t, r, w, \theta\}$ are the usual cylindrical coordinates, and the
hypersurfaces $\theta = 0, 2\pi$ are identified. The two Killing vectors are given by
$\xi_{(w)} = \partial_{w}$ and $\xi_{(\theta)} = \partial_{\theta}$.
To have cylindrical symmetry, some physical and geometrical conditions needed to be
imposed. In general this is not trivial. As a matter of fact,
when the symmetry axis is singular, it is still an open question: which
conditions  should be imposed \cite{Barnes}. Since in this paper we are mainly interested
in gravitational collapse, we would like to have the axis regular
in the beginning of the collapse. By this way, we are sure that the singularity
to be formed later on the axis is indeed due to the collapse. Thus, following
\cite{Fatima} we impose the following conditions:

(i) {\em There must exist a symmetry axis}. This can be written  as
\bq
\lb{cd1}
X \equiv \left|\xi^{\mu}_{(\theta)}\xi^{\nu}_{(\theta)}g_{\mu\nu}
\right|  \rightarrow 0,
\eq
as $r \rightarrow 0^{+}$, where we have chosen the radial coordinate $r$ such
that the axis  is located at $r = 0$.

(ii) {\em The spacetime near the symmetry axis is locally flat}. This can be
expressed as \cite{Kramer80}
\bq
\lb{cd2}
\frac{X_{,\alpha}X_{,\beta} g^{\alpha\beta}}{4X} \rightarrow - 1,
\eq
as  $r \rightarrow 0^{+}$, where $(\;)_{,\alpha} \equiv \partial (\;)/\partial
x^{\alpha}$. Note that solutions failing to satisfy this condition are sometimes
acceptable, and are usually expected that the singularities located on the axis
should be replaced by some
kind of sources in more realistic models. A particular case of these is
when the right-hand side of the above equation approaches a finite constant, and
the singularity now  can  be related to a line-like source
 \cite{VS}.  In this paper, since  we are mainly interested in
gravitational collapse, we shall not consider these possibilities and
assume that the above condition holds strictly at the initial of the collapse.

(iii) {\em No closed timelike curves (CTC's)}. In spacetimes with cylindrical symmetry,
CTC's can be easily introduced. To guarantee their absence,
 we impose the condition
\bq
\lb{cd3}
\xi^{\mu}_{(\theta)}\xi^{\nu}_{(\theta)}g_{\mu\nu} < 0,
\eq
in the whole spacetime.

In addition to these conditions, it is usually
also required that the spacetime be asymptotically flat in the radial direction.
However, since we consider solutions with self-similarity, this condition
cannot be satisfied by such solutions, unless we restrict the validity of
them only up to a maximal radius, say, $r = r_{0}(t)$, and
then join the solutions
with others in the region $r > r_{0}(t)$, which are
asymptotically flat in the radial direction.
In this paper, we shall not consider such a possibility, and simply assume
that  the self-similar solutions are valid in the whole spacetime.

Spacetimes with {\em homothetic self-similarity}
(or {\em self-similarity of the first kind})
is usually defined by the existence of a conform Killing vector $\xi^{\mu}$ that
satisfies the equations \cite{CT71},
\bq
\lb{2.3a}
\xi_{\mu;\nu} + \xi_{\nu;\mu} = 2 g_{\mu\nu},
\eq
where a semicolon ``;" denotes the covariant derivative. It can be shown that
for the spacetimes given by Eq.(\ref{2.1}) the conditions (\ref{2.3a}) imply
that
\bq
\lb{2.2}
M(t,r) = M(z),\;\;\;
S(t,r) = S(z),\;\;\;
V(t,r) = V(z),
\eq
where the self-similar variable $z$ and the corresponding conform Killing vector
$\xi^{\mu}$ are given by
\bq
\lb{2.3}
\xi^{\mu} \partial_{\mu} = t\partial_{t}
+ r \partial_{r},\;\;\;\;
z = \frac{r}{- t}.
\eq

It is interesting to note that under the coordinate transformations
\bq
\lb{res}
t = a_{1}\bar{t} + a_{2}\bar{r}, \;\;\;\;
r = a_{3} \bar{t} + a_{4}\bar{r},
\eq
the metric (\ref{2.1}), the regular conditions
(\ref{cd1})-(\ref{cd3}), and the self-similar
conditions (\ref{2.2}) and (\ref{2.3}) are all invariant,
where $a_{i}$'s are real constants, subject to $a_{1}a_{2} -  a_{3}a_{4} = 0$.
Using  this gauge freedom, we shall  assume that
\bq
\lb{cd4}
M(t, 0) = 0,
\eq
that is, the timelike coordinate $t$ measures the proper time on the axis.

\section{Self-similar Solutions of Massless Scalar Field }

\lb{SecIII}
\renewcommand{\theequation}{3.\arabic{equation}}
\setcounter{equation}{0}

For a massless scalar field, the Einstein field equations read
\bq
\lb{3.1}
R_{\mu\nu} = \kappa \phi_{,\mu}\phi_{,\nu},
\eq
where $\kappa [\equiv 8\pi G/c^{4}]$ is the  Einstein coupling  constant.
In this paper we shall choose units such that $\kappa  = 1$.
The scalar field satisfies the Klein-Gordon equation,
\bq
\lb{3.2}
g^{\alpha\beta}\phi_{;\alpha\beta} = 0.
\eq
However, this equation is not independent of the Einstein field equations (\ref{3.1}) and
can be obtained from the Biachi identities $G_{\mu\alpha;\beta}g^{\alpha\beta} = 0$.

On the other hand, it can be shown that a massless scalar field $\phi(t,r)$ that
is consistent with spacetimes with homothetic self-similarity must  take the form,
\bq
\lb{3.3}
\phi(t, r) = 2q\ln(-t) + \varphi(z),
\eq
where $q$ is an arbitrary constant, and $\varphi(z)$ is a function of $z$ only, which
will  be determined by the Einstein field equations (\ref{3.1}). Inserting Eqs.(\ref{A.3})
and (\ref{3.3}) into Eq.(\ref{3.1}) and considering the self-similar conditions
(\ref{2.2}), we find the following solutions
\bqn
\lb{3.4}
M(z) &=& 2 q^{2}\ln\left(1 - z^{2}\right),\;\;\;
S(z) = \ln(z),\nb\\
V(z) &=& -\ln(z),\;\;\;
\varphi(z) = 0.
\eqn
When $q = 0$ the corresponding spacetime is flat. Thus,
in the following we shall assume that $q \not= 0$.
Then, it can be shown that these solutions  satisfy all the conditions (\ref{cd1})-(\ref{cd3})
and (\ref{cd4}), and the corresponding Ricci and Kretschmann scalars are given by
\bqn
\lb{3.5}
R &=& g^{\alpha\beta}\phi_{,\alpha}\phi_{,\beta} = 4q^{2}
\frac{\left(1 - z^{2}\right)^{2q^{2}}}{t^{2}},\nb\\
I &\equiv& R^{\alpha\beta\lambda\sigma}R_{\alpha\beta\lambda\sigma}
= 48q^{4}
\frac{\left(1 - z^{2}\right)^{4q^{2}}}{t^{4}}.
\eqn
From these expressions we can see that the spacetime is singular on the hypersurface $t = 0$.
On the other hand, although the metric is singular on   $z = 1$, the
spacetime is not. Thus, to have a geodesically complete spacetime, we need to extend
the metric beyond this surface.    In order to do so, it is found convenient to
study the two cases $ 0 < 2 q^{2} < 1$ and $  2 q^{2} \ge 1$ separately.

\subsection{$ 0 < 2 q^{2} < 1$}

In this case, introducing two null coordinates $u$ and $v$ via the relations
\bqn
\lb{3.6}
t &=& -\left[(-u)^{n} + (-v)^{n}\right] \equiv - f_{+}(u,v),\nb\\
r &=& (-u)^{n} - (-v)^{n} \equiv  f_{-}(u,v),
\eqn
we find that in terms of $u$ and $v$ the metric and massless scalar field take the form
\bqn
\lb{3.7}
ds^{2} &=& n^{2}4^{1/n}f^{2(n-1)/n}_{+} dudv - f^{2}_{+}dw^{2}
           - f^{2}_{-}d\theta^{2},\nb\\
\phi &=& 2q\ln\left[f_{+}(u,v)\right],
\eqn
where
\bq
\lb{3.8}
n \equiv \frac{1}{1 - 2q^{2}} > 1.
\eq
From Eq.(\ref{3.6}) we can see that the region $t \le 0, \; r \ge 0, \; z < 1$
in the ($t,r$)-plane  is mapped into the region $u , \; v < 0,\; v \ge u$, which will
be referred to as Region $II$ [cf. Fig. 1]. The half line $z = 1,\; t \le 0$ is mapped
to $v = 0, \; u \le 0$. The region $v > 0,\; u \le 0$, which will be referred to as Region
$I$, is an extended region. Depending on the values of $n$, the nature of the
extension is different. In particular, it is analytical only for the case where $n$ is
an integer. Otherwise, the extension is not analytical, and
in some cases the metric and the scalar field
even become not real in this extended region, as we can see from Eqs.(\ref{3.6}) and
(\ref{3.7}).
To have the extension unique, in the following we shall consider only
 analytical extensions, that is,  the cases where $n$ is an integer.
Then, from Eqs.(\ref{3.7}) and (\ref{B.6}) we find that
\bqn
\lb{3.9}
\phi_{,u} &=& - 2nq\frac{ (-u)^{n-1}}{f_{+}},\;\;\;
\phi_{,v} = - 2nq\frac{(-v)^{n-1}}{f_{+}},\nb\\
R &=& \phi_{,\alpha} \phi^{,\alpha} = n^{4}q^{2}4^{1 + 1/n}
\frac{(uv)^{n-1}}{{f_{+}}^{2/n}},\nb\\
I &=& R^{\alpha\beta\lambda\sigma}R_{\alpha\beta\lambda\sigma}
    = 16^{2-1/n}\left(\frac{n-1}{n}\right)^{2}
    \frac{(uv)^{2(n-1)}}{{f_{+}}^{4(2-1/n)}},\nb\\
\Theta_{l} &=&  \frac{4^{1-1/n}(-v)^{2n -1}}{nf_{-}f_{+}^{2(2n-1)/n}},\;\;\;
\Theta_{n} = -  \frac{4^{1-1/n}(-u)^{2n -1}}{nf_{-}f_{+}^{2(2n-1)/n}}.
\eqn
From these expressions we can see that the spacetime is regular on the symmetry axis
$v = u$ in Region $II$, and $\phi_{,\alpha}$ is always timelike. On the hypersurface
$v = 0$, we have
\bq
\lb{3.10a}
\phi_{,v}(u, 0) = 0,
\eq
and the only non-vanishing component of
the energy-momentum tensor $T_{\mu\nu}$ is given by
\bq
\lb{3.10}
T_{uu}(u, 0) = \phi_{,u}^{2}(u, 0) \not= 0,
\eq
which represents an energy flow, moving from Region $II$ into Region $I$ along the
null hypersurfaces $u = Const.$ The expansion, $\Theta_{l}$, of
the null geodesics along the hypersurfaces $u =  Const.$ is always positive
in this region, while the expansion, $\Theta_{n}$, of
the null geodesics along the hypersurfaces $v =  Const.$ is always negative.
However, $\Theta_{l}$ becomes zero on the hypersurface $v = 0$ and then negative
in the extended region, $I$, where $v > 0$, while $\Theta_{n}$ is negative even in this
extended region. Thus, all the cylinders of constant $t$ and $r$ are trapped in
the extended region, but not in Region $II$.
Then, the hypersurface $v = 0$ defines an apparent horizon  \cite{HE73,Hay94}.

It should be noted that the above analysis is very important when we consider boundary
conditions on the apparent horizon in the next section, as it shows
clearly that it is the component $\phi_{,u}$  that represents the energy flow of
the scalar waves  that moves from   Region $II$ into Region $I$, while the component
$\phi_{,v}$  represents the energy flow of the scalar field that moves in the opposite
direction. Since now Region $I$ is a trapped region and no radiation
is able to escape from this region. This can be seen clearly from Eq.(\ref{3.10a}).

The singularity
behavior in Region $I$ depends on the values of $n$.
In particular, when $n$ is an odd integer,
from Eq.(\ref{3.9}) we can see that the spacetime becomes singular on the hypersurface
$r = 0$ or $u = -v$, which services as
the up boundary of the spacetime,
and the corresponding Penrose diagram is that of Fig. 1. Thus,
in this case Region $I$ can be
considered as the interior of a black hole, and the corresponding
solutions   represent gravitational collapse of a massless scalar
field in Region $II$. The collapse  always  forms  a black hole.
Note that in this case $\phi_{,\alpha}$  is continuously  timelike in the
trapped region, $I$,  as we can see from Eq.(\ref{3.9}).

%%%%%%%%%%%%%%%%%%%%%%%%%%%%%%%%%%%%%%%%%%%%%%%%%%%%%%%%%%%%%%%%%%%%%%%%%%%%
%%
 \begin{figure}[htbp]
 \begin{center}
 \label{fig1}
 \leavevmode
  \epsfig{file=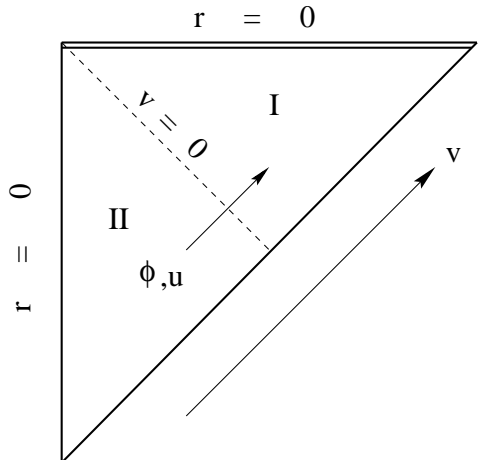,width=0.35\textwidth,angle=0}
 \caption{The Penrose diagram for the solutions given  by
 Eqs. (\ref{3.6}) and (\ref{3.7}) with $n \ge 2$ being an integer.
 The cylinders of constant $t$ and $r$
are all trapped in Region $I$ where $\Theta_{l} < 0$
and $\Theta_{l}\Theta_{n} > 0$,
but not in Region $II$, where $\Theta_{l} > 0$ and
$\Theta_{l}\Theta_{n} < 0$. The dashed
line $v = 0$ represents the apparent horizon.
When $n$ is an odd integer, the spacetime is singular on the horizontal double
line $r = 0$, and when $n$ is an even integer, the spacetime has no curvature
singularity there.}
 \end{center}
 \end{figure}
%%%%%%%%%%%%%%%%%%%%%%%%%%%%%%%%%%%%%%%%%%%%%%%%%%%%%%%%%%%%%%%%%%%%%%%%%%%%
%%

 When $n$ is an even integer,  from Eq.(\ref{3.9})
 we can see that the Ricci and Kretschmann
scalars are all finite at $r = 0$ in
Region $I$, but both of $\Theta_{l}$
and $\Theta_{n}$ become singular there. Thus,
anything that moves along the null geodesics,
defined by $l^{\mu}$ or $n^{\mu}$,  will be crashed
to zero volume by the infinitely large contraction.
Then, the hypersurface $r = 0$ now represents a
topological boundary of the spacetime, and the
corresponding Penrose diagram is also given by Fig. 1,
but now the spacetime is free of curvature
singularities on the double horizontal line $ r = 0$.

It should be noted that apparent horizons and black holes are usually defined
in asymptotically flat spacetimes \cite{HE73}. To be distinguishable,
Hayward called such apparent horizons as trapping
horizons and defined black holes by the  future outer trapping horizons
\cite{Hay94}. For the sake of simplicity and without causing any confusions,
in this paper we shall continuously use the notions of apparent horizons
in the places of Hayward's trapping horizons, and define black holes in
a little bit more general sense than that of Hayward in non-asymptotically
flat spacetimes.

\subsection{$2 q^{2} \ge 1$}

In this case, introducing the two null coordinates $u$ and $v$ via the relations
\bq
\lb{3.11}
t = u + v, \;\;\; r  = v - u,
\eq
we find that the metric and scalar field are given by
\bqn
\lb{3.12}
ds^{2} &=& 4^{1-2q^{2}}\left[\frac{(u+v)^{2}}{uv}\right]^{2q^{2}} du dv
- (u + v)^{2} dw^{2} - (u - v)^{2}d\theta^{2},\nb\\
\phi &=& 2q\ln\left[-(u + v)\right].
\eqn
To study the physics of the spacetime near the hypersurface $v = 0$ or $z = 1$
in some details, let us consider the radial null geodesics along the hypersurface
$u = Const. $, say, $ u = u_{0}$,
\bq
\lb{3.13}
\ddot{v} - 2q^{2}\frac{u_{0} - v}{v(u_{0} + v)} \dot{v}^{2} = 0,
\eq
where an over-dot denotes the ordinary differentiation with respect to the affine
parameter $\lambda$ along the null geodesics. Then, near the hypersurface $v = 0$,
Eq.(\ref{3.13}) has the solution
\bq
\lb{3.14}
v(\lambda) = \cases{\left(b_{1}\lambda + b_{2}\right)^{-1/(2q^{2}-1)},
            & $2q^{2} > 1$,\cr
e^{b_{1}\lambda + b_{2}}, & $ 2q^{2} = 1$,\cr}
\eq
where $b_{1}$ and $b_{2}$ are the integration constants.
Thus, as $v \rightarrow 0$, we must have $\lambda \rightarrow \pm \infty$. That is, the
``distance" between the point $(u_{0}, 0)$ and any of the other points, say,
$(u, v) = (u_{0}, v_{0} < 0)$,  along the null geodesics $u = u_{0}$ is infinite.
Therefore, when $2 q^{2} \ge 1$  the hypersurface
$v = 0$ actually represents a natural boundary of the spacetime,
and there is no need to extend
the solutions beyond this surface, since now Region $II$ is already geodesically maximal.
It should be noted that, although there is no spacetime singularity  on the
half-line $v = 0, u < 0$,
the spacetime is singular at the point $(u, v) = (0,0)$, as can be seen from
Eq.(\ref{3.9}). The corresponding Penrose diagram is given by Fig. 2.

%%%%%%%%%%%%%%%%%%%%%%%%%%%%%%%%%%%%%%%%%%%%%%%%%%%%%%%%%%%%%%%%%%%%%%%%%%%%
%%
 \begin{figure}[htbp]
 \begin{center}
 \label{fig2}
 \leavevmode
  \epsfig{file=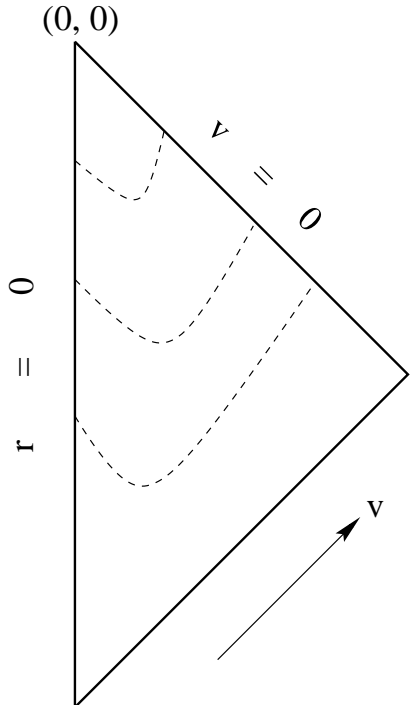,width=0.3\textwidth,angle=0}
 \caption{The Penrose diagram for the solutions given  by
 Eq. (\ref{3.12}) for $2 q^{2} \ge 1$. The spacetime is geodesically
  maximal in the whole region $u , v < 0,\; v \ge u$,
   and singular only at $(u,v) = (0,0)$.
 The two surfaces of constant $t$ and $r$ are   not trapped,
 because  now we always have $\Theta_{l} > 0$ and $\Theta_{n} < 0$. The
 only exception is on the surface $v = 0, \; u < 0$ where $\Theta_{l}(u, 0) = 0$
 and $\Theta_{n}(u, 0) < 0$. The dashed
  lines   represent the hypersurfaces $\phi(u,v) = Const.$,
  which are always spacelike.}
 \end{center}
 \end{figure}
%%%%%%%%%%%%%%%%%%%%%%%%%%%%%%%%%%%%%%%%%%%%%%%%%%%%%%%%%%%%%%%%%%%%%%%%%%%%
%%

On the other hand, from Eq.(\ref{B.6}) we find that
\bqn
\lb{3.15}
\Theta_{l} & =& - 4^{1-2q^{2}}\left[\frac{(u+v)^{2}}{uv}\right]^{2q^{2}}
\frac{v}{u^{2} - v^{2}},\nb\\
\Theta_{n} & =&  4^{1-2q^{2}}\left[\frac{(u+v)^{2}}{uv}\right]^{2q^{2}}
\frac{u}{u^{2} - v^{2}}.
\eqn
Thus, in the whole spacetime now we always have
$\Theta_{l} > 0$ and $\Theta_{n} < 0$,
except on the half-hypersurface $v = 0, \; u < 0$
where we have $\Theta_{l} = 0$, $\; \Theta_{n} < 0$. That is,
all the two surfaces of constant $t$ and $r$ are not trapped for $u, v < 0$, and
become    marginally trapped only on the half surface $v = 0, u < 0$.
In addition, we have
\bq
\lb{3.16}
R = \phi_{,\alpha} \phi^{,\alpha} =  4^{1+2q^{2}}q^{2}
\frac{(uv)^{2q^{2}}}{(u+v)^{2(1+ 2q^{2})}},
\eq
which is always positive for $u, v < 0$, and zero only when $v = 0$.
That is, the scalar field is always timelike, except
on the hypersurface $v = 0$ where it becomes null.
Thus, in this case the corresponding
solution can be considered as representing gravitational
collapse of a massless scalar field. Although now no black holes are formed, a
point-like spacetime singularity is indeed developed at the point
$(u, v) = (0, 0)$, as we can see from Eq.(\ref{3.5}). It is interesting to
note that this singularity is not naked, and an observer can see it
only when he/she arrives at that point.

\section{Linear Perturbations of the Self-Similar Solutions}
\lb{SecIV}
\renewcommand{\theequation}{4.\arabic{equation}}
\setcounter{equation}{0}

To see if the above solutions represent critical collapse, we need to do
their linear perturbations, because by definition a critical solution has
one and only one unstable mode. To study such perturbations, it is found convenient
to use the self-similar variables $\tau$ and $z$ defined by Eq.(\ref{A.2}) but
still work in the ($t,r$)-coordinates.  Then, the
 linear perturbations can be written as
\bq \lb{4.1} F(\tau, z) = F_{0}(z) + \epsilon F_{1}(z) e^{k \tau},
\eq where $F \equiv \{M, S, V, \varphi\}$, and $\epsilon$ is a
very small real constant. Quantities with subscripts ``1" denote
perturbations, and those with ``0" denote the background
self-similar solutions given by Eq.(\ref{3.4}). It is understood
that there may be many perturbation modes for different values
(possibly complex) of the constant $k$. Then, the general
perturbations will be the sum of these individual ones. Modes with
$Re(k) > 0$ grow as $\tau \rightarrow \infty$ and are referred to
as unstable modes, and the ones with $Re(k) < 0$ decay and are
referred to as stable modes.

It should be noted that in writing Eq.(\ref{4.1}), we have already used some
of the gauge freedom to write the perturbations
such that they preserve the form of the metric (\ref{2.1}).
However, this does not completely fix the gauge.
We shall return to this point later when we consider the gauge modes.

To the first order of $\epsilon$, the Ricci tensor is given by
Eqs.(\ref{A.5a})-(\ref{A.6b}). Applying them to the background solutions given
by Eq.(\ref{3.4}), and using the Einstein field equations (\ref{3.1}), we find
that there are only four independent equations, which can be cast in the form,
\bqn
\lb{4.2}
k M_{1}(z) &=& 2z^{2}S''_{1} + 2z V'_{1}
              + 2z\left[(1 + k) - 4q^{2}\frac{z^{2}}{1-z^{2}}\right]S'_{1}
              + kV_{1}\nb\\
           & & + k\left(1 - 4q^{2}\frac{z^{2}}{1-z^{2}}\right)S_{1}
               + 4qz\varphi'_{1},
\eqn
and
\bqn
\lb{4.3a}
z\left(1-z^{2}\right)V''_{1} + \left[1 - (1 + 2k)z^{2}\right]V'_{1}
   - k^{2}zV_{1} &=& kzS_{1} - \left(1-z^{2}\right)S'_{1},\\
\lb{4.3b}
z\left(1-z^{2}\right)\varphi''_{1} + \left[1 - (1 + 2k)z^{2}\right]\varphi'_{1}
   - k^{2}z\varphi_{1} &=& 2qz\left(zS'_{1} + kS_{1}\right),\\
\lb{4.3c}
z\left(1-z^{2}\right)S''_{1} + 2\left(1 -kz^{2}\right)S'_{1}
   +k(1-k) zS_{1} &=& 0,
\eqn
where a prime denotes the ordinary differentiation with respect to the indicated
argument. It can be shown that Eq.(\ref{4.3c}) has the general solution,
\bq
\lb{4.4}
S_{1}(z) = \frac{1}{z}\left[c_{1}(1+z)^{2-k} + c_{2}(1-z)^{2-k}\right],
\eq
where $c_{1}$ and $c_{2}$ are two integration constants. Substituting the above
solution into Eqs.(\ref{4.3a}) and (\ref{4.3b}) we find that these two equations
can be written in the form
\bq
\lb{4.5}
y(1-y)\frac{d^{2}Z_{i}}{dy^{2}} + \left[e - (a + b+1)y\right] \frac{dZ_{i}}{dy}
- ab Z_{i} = \frac{1}{4y^{1/2}} f_{i}(y),
\eq
where $y \equiv z^{2}$, $\;\left\{Z_{i}\right\} =\left\{V_{1}, \varphi_{1}\right\}$,
$\; a = b = k/2, \;  e = 1$, and
\bqn
\lb{4.6}
f_{1}(y) &\equiv& kzS_{1} - \left(1-z^{2}\right)S'_{1}\nb\\
       &=& \frac{1}{y}\left\{c_{1}\left[1 - (2-k)y^{1/2} + y\right]
       \left(1+y^{1/2}\right)^{2-k}
           + c_{2}\left[1 + (2-k)y^{1/2} + y\right]
           \left(1-y^{1/2}\right)^{2-k}\right\},\nb\\
f_{2}(z) &\equiv& 2qz\left(zS'_{1} + kS_{1}\right)\nb\\
       &=& -2q\left\{c_{1}\left[(1-k) -y^{1/2}\right]
       \left(1+y^{1/2}\right)^{1-k}
           + c_{2}\left[(1-k) + y^{1/2}\right]
           \left(1-y^{1/2}\right)^{1-k}\right\}.
\eqn
Eq.(\ref{4.5}) is the inhomogeneous hypergeometric equation \cite{Bateman}, and the
general solution of the associated homogeneous  equation is a linear combination
of the two independent solutions, $F^{(i)}_{1}(z)$, where
\bqn
\lb{4.7}
F_{1}^{(1)}(z) &=& F\left(\frac{1}{2}k, \frac{1}{2}k; 1; z^{2}\right),\nb\\
F_{1}^{(2)}(z) &=& F\left(\frac{1}{2}k, \frac{1}{2}k; k; 1-z^{2}\right),
\eqn
with $F(a, b; e; z)$ denoting the hypergeometric function. From the above two
independent solutions, we can construct particular solutions of the inhomogeneous
equation (\ref{4.5}), and then find that the general solutions for $V_{1}(z)$ and
$\varphi_{1}(z)$ can be written as
\bqn
\lb{4.8}
V_{1}(z) &=& \left(a^{(2)}_{1} + A^{(2)}_{1}(z)\right) F^{(1)}_{1}(z)
              + \left(a^{(1)}_{1} - A^{(1)}_{1}(z)\right) F^{(2)}_{1}(z),\nb\\
\varphi_{1}(z) &=& \left(a^{(2)}_{2} + A^{(2)}_{2}(z)\right) F^{(1)}_{1}(z)
              + \left(a^{(1)}_{2} - A^{(1)}_{2}(z)\right) F^{(2)}_{1}(z),
\eqn
where $a^{(i)}_{j}$'s are  integration constants, and
\bqn
\lb{4.9}
A^{(i)}_{j}(z) &\equiv& \int^{z}{\frac{f_{j}(z)F^{(i)}_{1}(z) dz}
{z\left(1 - z^{2}\right)\Delta(z)}},\nb\\
\Delta(z) &\equiv& F^{(2)}_{1}(z)\frac{d}{dz}\left(F^{(1)}_{1}(z)\right)
   -  F^{(1)}_{1}(z)\frac{d}{dz}\left(F^{(2)}_{1}(z)\right).
\eqn

To have physically acceptable perturbations, we need to impose
boundary conditions. In General Relativity, this is a very subtle problem and
there are no fixed rules to follow. In this paper we shall choose the axis
$r = 0$ and the hypersurface $z = 1$ as the places where we impose the boundary
conditions.

Since the axis for the background solutions is regular, and the conditions
(\ref{cd1})-(\ref{cd2}) and (\ref{cd4}) are satisfied by them, we would
expect that the linear perturbations also satisfy these conditions. In particular,
it can be shown that the condition (\ref{cd4}) requires $M_{1}(0) = 0$,
and the ones (\ref{cd1}) and (\ref{cd2}) require, respectively,
$G_{1}(z) \rightarrow 0$ and $z{G'}_{1}(z)\rightarrow 0$,
as $z \rightarrow 0$, where $G_{1}(z) \equiv S_{1}(z) + V_{1}(z)$.
On the other hand, the condition that $R = \phi_{,\alpha}\phi^{,\alpha}$
is regular on the axis further requires $z{\varphi'}_{1} + k \varphi_{1}
\rightarrow 0$ \footnote{It should be noted that this
condition is not independent of the ones (\ref{cd1}) and (\ref{cd2}).
In fact, using the Einstein field equations (\ref{3.1}) we can deduce it
from Eqs.(\ref{cd1}) and (\ref{cd2}). However, without loss of generality,
in this paper we shall impose it   independently.}. In summary, on the
axis we shall impose the following conditions,
\bqn
\lb{cd5}
& & M_{1}(z) \rightarrow 0, \nb\\
& & G_{1}(z) \rightarrow 0,\nb\\
& & z{G'}_{1}(z)  \rightarrow 0, \nb\\
& & z{\varphi'}_{1}(z) + k \varphi_{1}(z) \rightarrow 0,
\eqn
as $z \rightarrow 0$.

On the other hand, when $0 < 2q^{2} < 1$ the hypersurface $z = 1$ is an apparent
horizon [cf. Fig.1 ], and  we required that the background solutions
be analytical across it with respect to the null coordinate $v$.
Otherwise, it was found that the extension was not
unique. Clearly, this condition should hold also for the perturbations.
In addition, since the hypersurface $z=1$ represent an apparent horizon
and Region $I$ is a trapped region, so nothing should be able to escape from it.
In particular, for the scalar field this implies $\phi_{,v}(u, 0) = 0$
[cf. Eq.(\ref{3.10})].   On the other
hand, as shown in \cite{TW90}, the gravitational wave component that moves out
of  Region $I$ is represented by $\Psi_{0}$, which is a function
of $V_{,vv}$, $\; V_{,v}$, $\; M_{,v}$ and $S_{,v}$.
%%%%%%%%%%%%%%%%%%%%%%%%%%%%%%%%%%%%%%%%%%%%%%%%%%%%%%%%%%%%%%%%%%%%%%%%%%%%%%%%%%
%given by
%\bq
%\lb{4.10a}
%\Psi_{0} \equiv - C_{\alpha\beta\gamma\delta}l^{\alpha}m^{\beta}l^{\gamma}
%m^{\delta} = - \frac{1}{2}e^{-2\sigma} \left[V_{,vv}
%- \left(2\sigma_{,v} - U_{,v}\right)V_{,v}\right],
%\eq
%where $v$ is the null coordinate defined by Eq.(\ref{3.6}), $U \equiv S - 2\ln(r)$,
%and $\sigma [\equiv \frac{1}{2}\ln(g_{uv})]$ can be read off from Eq.(\ref{3.7}).
%$m^{\mu}$ is a complex vector defined by Eq.(3) in \cite{TW90} and
%$C_{\alpha\beta\gamma\delta}$ denotes the Weyl tensor. For a detail,
%we refer readers to \cite{TW90}.
%%%%%%%%%%%%%%%%%%%%%%%%%%%%%%%%%%%%%%%%%%%%%%%%%%%%%%%%%%%%%%%%%%%%%%%%%%%%%%%%%%%
Then, we can see that the condition that no gravitational waves come
out from Region $I$  requires $V_{,vv}$, $\; V_{,v}$, $\; S_{,v}$,
$\; M_{,v}$, $\; \phi_{,v} \rightarrow 0$
 as $v \rightarrow 0$. Changing to the self-similar variable
$z$, it can be shown that these  conditions are equivalent to
\bqn
\lb{cd6}
& & F_{1}(z) \sim {\mbox{ analytical with respect to $v$}}, \nb\\
& & (1-z)^{(n-1)/n}\frac{dF_{1}(z)}{dz} \rightarrow 0, \\
& & (1-z)^{(n-2)/n}\frac{dV_{1}(z)}{dz} \rightarrow 0,  \;\;\;\; (0 < 2q^{2} < 1),\nb
\eqn
as $z \rightarrow 1$, where $F_{1}(z) \equiv \left\{M_{1}, S_{1},
V_{1}, \varphi_{1}\right\}$.

When $2q^{2} \ge 1$ the hypersurface $z = 1$ is a future null infinity
of the spacetime and marginally trapped [cf. Fig. 2]. Since it is the future null
infinity, we would expect that the perturbations be finite there. On the other hand,
the hypersurface $z =1$ is also marginally trapped, so there should have only
outgoing scalar field and gravitational waves. Note that the amplitude of
gravitational wave components is always proportional to $e^{M_{0}} \sim
(1-z)^{2q^{2}}$ \cite{TW90}. Thus, now we must require
$(1-z)^{2q^{2}}(V_{1,vv}$, $\; V_{1,v} $, $\; S_{1,v} $,
$\; M_{1,v})$ $\; \rightarrow 0$, as $z \rightarrow 1$.
On the other hand, using the Einstein field equations we can show that these
conditions also imply $(1-z)^{2q^{2}}\varphi_{1,v}$ $\; \rightarrow 0$.
In terms of the self-similar variable $z$, and
noticing that now $t$ and $r$ are given by
Eq.(\ref{3.11}), we find that these conditions can be written as
\bqn
\lb{cd7}
&& F_{1}(z) \rightarrow {\mbox{finite}},\nb\\
& & (1-z)^{2q^{2}} \frac{dF_{1}(z)}{dz} \rightarrow 0, \nb\\
& & (1-z)^{2q^{2}} \frac{d^{2}V_{1}(z)}{dz^{2}} \rightarrow 0,\;\;
(2q^{2} \ge 1),
\eqn
as $z \rightarrow 1$.
%%%%%%%%%%%%%%%%%%%%%%%%%%%%%%%%%%%%%%%%%%%%%%%%%%%%%%%%%%%%%%%%%%%%%%%%%%
%%Note that in this case it seems that there is no reason to
%%require that the metric coefficients be analytical as $v \rightarrow 0$.
%%%%%%%%%%%%%%%%%%%%%%%%%%%%%%%%%%%%%%%%%%%%%%%%%%%%%%%%%%%%%%%%%%%%%%%%%%

Once we have the boundary conditions, let us first consider the gauge
modes. We note that the metric (\ref{2.1}) is invariant under the
coordinate transformations,
\bqn
\lb{eq1}
t &=& a(\bar{t} + \bar{r}) + b(\bar{t} - \bar{r}),\nb\\
r &=& a(\bar{t} + \bar{r}) - b(\bar{t} - \bar{r}),
\eqn
where $a(\bar{t} + \bar{r})$ and  $b(\bar{t} - \bar{r})$ are arbitrary
functions of their indicated
arguments, subject to $a'b' \not= 0$. Thus,
let us consider the gauge transformations
\bqn
\lb{eq2}
t &\rightarrow& t + \epsilon\left[A(t + r) + B(t - r)\right],\nb\\
r &\rightarrow& r + \epsilon\left[A(t + r) - B(t - r)\right],
\eqn
where $A(t + r)$ and  $B(t - r)$ are  other arbitrary functions. Then, we
find that under the above coordinate transformations the resultant perturbations
are given by
\bq
\lb{eq3}
F(\tau, z) = F_{0}(z) + {\delta}F(\tau, z),
\eq
with
\bqn
\lb{eq4}
{\delta}M(\tau, z) &=& 2(A' + B') - \frac{4q^{2}z}{t(1-z^{2})}
\left[(A - B) + (A +B)z\right],\nb\\
{\delta}S(\tau, z) &=& \frac{1}{r}
\left[(A - B) + (A +B)z\right],\nb\\
{\delta}V(\tau, z) &=& - \frac{1}{r}
\left[(A - B) + (A +B)z\right],\nb\\
{\delta}\varphi(\tau, z) &=&  \frac{4q }{t}(A +B).
\eqn
In order to have the above expressions be in the form of Eq.(\ref{4.1}), we must
choose
\bqn
\lb{eq5}
A(t + r) &=& - c_{2}t_{0}\left(\frac{-t}{t_{0}}\right)^{1-k}(1-z)^{1-k},\nb\\
B(t - r) &=& c_{1}t_{0}\left(\frac{-t}{t_{0}}\right)^{1-k}(1+z)^{1-k},
\eqn
for which   Eq.(\ref{eq4}) can be written as
\bq
\lb{eq6}
{\delta}F(\tau, z) = F_{1}(z)e^{k\tau},
\eq
with
\bqn
\lb{gaugemode}
M_{1}(z) &=& 2\left\{c_{1}\left[(1-k) + 2q^{2}z\right](1+z)^{-k}
- c_{2}\left[(1-k) - 2q^{2}z\right](1- z)^{-k}\right\},\nb\\
S_{1}(z) &=& \frac{1}{z} \left[c_{1}(1+z)^{2-k}
    + c_{2} (1- z)^{2-k}\right],\nb\\
V_{1}(z) &=& \frac{1- z^{2}}{z} \left[c_{1}(1+z)^{-k}
    + c_{2} (1- z)^{-k}\right],\nb\\
\varphi_{1}(z) &=& -2q   \left[c_{1}(1+z)^{1-k}
    - c_{2} (1- z)^{1-k}\right],
\eqn
where $c_{1}$ and $c_{2}$ are arbitrary constants. It can be shown that the above
perturbations don't satisfy the boundary conditions (\ref{cd5})-(\ref{cd7}), or
in other words, the boundary conditions imposed in this paper limit all the gauge
modes.

Now let us  consider the above boundary conditions
for the particular background  given by Eq.(\ref{3.4}). We first consider the
conditions at the axis given by Eq.(\ref{cd5}). Let us first note that
\bqn
\lb{4.10}
F^{(1)}_{1}(z) &=& 1 + \frac{1}{4}k^{2}z^{2}
                    + O\left(z^{4}\right),\nb\\
 F^{(2)}_{1}(z) &=& A_{2}(k)
       \left\{\left[A_{1}(k) - 2\ln(z)\right]
       F^{(1)}_{1}(z) + \frac{1}{2}k(k-2)z^{2}\right\},
\eqn
as $z \rightarrow 0$, where
\bq
\lb{4.11}
A_{1}(k) = \sum^{\infty}_{n = 0}{\frac{2(2-k)}{(n + 1)(2n +k)}},
\;\;\;\;
A_{2} \equiv \frac{\Gamma(k)}{\Gamma^{2}\left(\frac{1}{2}k\right)},
\eq
and $\Gamma(k)$ denotes the gamma function. Inserting Eqs.(\ref{4.10})
into Eqs.(\ref{4.9}) and (\ref{4.8}), after  tedious calculations we find
that
\bq
\lb{4.12}
z{\varphi'}_{1}(z) + k \varphi_{1}(z) \rightarrow
   -2k a^{(1)}_{2}A_{2}(k) \ln(z) + A_{0}(k),
\eq
as $z \rightarrow 0$, where $A_{0}(k)$ is a finite constant. Thus, the last
condition of Eq.(\ref{cd5}) requires
\bq
\lb{4.13}
a^{(1)}_{2} = 0.
\eq
Similarly, one can show that
\bqn
\lb{4.14}
G_{1}(z) \equiv S_{1}(z) + V_{1}(z) &=& \frac{2\left(c_{1} + c_{2}\right)}{z}
           - 2a^{(1)}_{1}A_{2}(k) \ln(z) \nb\\
           & & + (2-k)\left(c_{1} - c_{2}\right)
           + \left(a^{(1)}_{1}A_{1}(k)A_{2}(k) + a^{(2)}_{1}\right)
           + O\left(z\right),
\eqn
from which we can see that
the second condition of Eq.(\ref{cd5}) requires
\bq
\lb{4.15}
c_{1} = - c_{2} = c,\;\;\;
a^{(1)}_{1} = 0,\;\;\;
a^{(2)}_{1} = 2(k-2)c.
\eq
Once Eq.(\ref{4.15}) holds, it can be shown that
\bqn
\lb{4.16}
z{G'}_{1}(z) &\rightarrow& O\left(z^{2}\right),\nb\\
M_{1}(z) &\rightarrow&   O\left(z^{2}\right),
\eqn
as $z \rightarrow 0$.  That is, the first and third conditions of
Eq.(\ref{cd5}) do not impose further restrictions on the free parameters.

Now let us turn to consider the boundary conditions at $z = 1$. It is found convenient
to consider the cases $0 < 2q^{2} <1$ and $2q^{2} \ge 1$ separately.

\subsection{$0 < 2q^{2} <1$}

In this case we have $1 - z \sim (-v)^{n}$, as $v \rightarrow 0$. Then, we find that
\bq
\lb{4.17}
S_{1}(z) \sim c \left(2^{2-k} - (-v)^{(2-k)n}\right),
\eq
as $v \rightarrow 0$. Thus, in order to have $S_{1}(z)$ be analytical,
the constant $k$ has to take the values,
\bq
\lb{4.18}
k = 2 - \frac{m}{n}, \;\;\; (m \ge 1),
\eq
where $m$ is a positive integer ($m = 1, 2, 3, ...)$. Since when $k < 0$ the corresponding
modes are stable, which we are not so interested in. Thus, in the following we shall consider
only the case where $k > 0$, which together with Eq.(\ref{4.18}) implies $2n > m \ge 1$.
To study the boundary conditions at $z = 1$ further, let us consider
the cases $k \not= 1$ and $k = 1$ separately.

{\bf Case A)} $\;  k \not= 1$: In this case it can be shown that
\bqn
\lb{4.19}
F^{(1)}_{1}(z) &=& B_{1}(k) F^{(1)}_{11}(z) + B_{2}(k)F^{(1)}_{12}(z)x^{1-k},\nb\\
F^{(2)}_{1}(z) &=& F^{(1)}_{11}(z),\;\;\; (k \not= 1),
\eqn
where
\bqn
\lb{4.20}
F^{(1)}_{11}(z) &\equiv & 1 + \frac{1}{4}k x + D_{1}(k) x^{2}
+ O\left(x^{3}\right),\nb\\
F^{(1)}_{12}(z) &\equiv & 1 + \frac{1}{4}(2-k)x + D_{2}(k) x^{2}
  + D_{3}(k) x^{3} + O\left(x^{4}\right),
\eqn
with $x \equiv 1 - z^{2}$ and
\bqn
\lb{4.21}
B_{1}(k) &=& \frac{\Gamma(1-k)}{\Gamma^{2}\left(1 - \frac{1}{2}k\right)},
\;\;\;\;
B_{2}(k) = \frac{\Gamma(k -1)}{\Gamma^{2}\left(\frac{1}{2}k\right)},\nb\\
D_{1}(k) &=& \frac{k(2+k)^{2}}{32(1+k)},\;\;\;
D_{2}(k) = \frac{(2-k)(4-k)^{2}}{32(3-k)},\nb\\
D_{3}(k) &=&   \frac{(2-k)(4-k)(6-k)^{2}}{384(3-k)}.
\eqn

Inserting the above expressions into Eqs.(\ref{4.2}), (\ref{4.8})
and (\ref{4.9}) and after tedious calculations we find that
\bqn
\lb{4.22a}
kM_{1}(z) &=& \frac{1}{2}k\left\{c(k-2)\left(k-1 - 2q^{2}\right)2^{k}
- 2B_{2}(k)\left[(k-1)a^{(2)}_{1} + 2kqa^{(2)}_{2}\right]\right\}x^{1-k}\nb\\
& & - 2\left\{c(k-2)\left(k-1 - 2q^{2}\right)2^{k}
+2(1-k)B_{2}(k)\left[a^{(2)}_{1} + 2qa^{(2)}_{2}\right]\right\}x^{-k}\nb\\
& & + A_{3}(c, k, q) + O\left(x^{k}\right) + O\left(x^{2-k}\right),\\
\lb{4.22b}
\varphi_{1}(z) &=&  a^{(2)}_{2} \left[B_{2}(k)x^{1-k} + B_{1}(k)\right]
  + O\left(x\right),\\
  \lb{4.22c}
V_{1}(z) &=&  a^{(2)}_{1} B_{2}(k)x^{1-k} + O\left(x^{2-k}\right),
\;\;\; (k \not= 1),
\eqn
as $z \rightarrow 1$, where $A_{3}(c, k, q)$ is a finite constant, and $a^{(2)}_{1}$
is given by Eq.(\ref{4.15}). From the last two equations we can see that the analytical
conditions of $V_{1}(z)$ and $\varphi_{1}(z)$ require $k < 1$, which
together with Eq.(\ref{4.18}) implies
\bq
\lb{4.23}
2n > m > n.
\eq
The condition that $M_{1}(z)$ is analytical across the hypersurface $z = 1$ requires
\bq
\lb{4.24}
a^{(2)}_{2} = \frac{c(k-2)}{2}\left(\frac{(k - 1 -2q^{2})2^{k}}{2(k-1)B_{2}(k)}
- 2\right).
\eq
On the other hand, it can be also shown that
\bqn
\lb{4.25}
x^{1 - 1/n}\frac{dM_{1}(x)}{dx} &\sim& \frac{c(k-2)}{2}
\left[\left(k - 1 -2q^{2}\right)2^{k} + 4(1-k)B_{2}(k)\right]x^{(m-n-1)/n}
+ O\left(x^{(n-1)/n}\right),\nb\\
x^{1 - 1/n}\frac{dV_{1}(x)}{dx} &\sim& (1-k)a^{(2)}_{1} B_{2}(k)x^{(m-n-2)/n}
+ O\left(x^{1-2/n}\right) + O\left(x^{(m-2)/n}\right),\nb\\
x^{1 - 1/n}\frac{d\varphi_{1}(x)}{dx} &\sim& (1-k)a^{(2)}_{2} B_{2}(k)x^{(m-n-1)/n}
+ O\left(x^{(n-1)/n}\right),\nb\\
x^{1 - 1/n}\frac{dS_{1}(x)}{dx} &\sim& \frac{ck}{2^{k}}  x^{(n-1)/n}
+ O\left(x^{(m-1)/n}\right), \;\;\; (k \not=1),
\eqn
as $x \rightarrow 0$. Thus, the last two conditions of Eq.(\ref{cd6}) further require
\bq
\lb{4.26}
2n > m > n + 2, \;\;\; (k \not=1).
\eq

{\bf Case B)} $\; k =1$: In this case it can be shown that
\bqn
\lb{4.27}
F^{(1)}_{1}(z) &=& F^{(1)}_{11}(x) - F^{(1)}_{12}(x) \ln(x),\nb\\
F^{(2)}_{1}(z) &=& \pi F^{(1)}_{12}(x),\;\;\; (k =1),
\eqn
but now with
\bqn
\lb{4.28}
F^{(1)}_{11}(x) &\equiv& \frac{2}{\pi}\left[C_{0} + \frac{1}{4}(C_{0} -1)x
+ \frac{3}{128}(6C_{0} - 7)x ^{2}
+  O\left(x^{3}\right)\right],\nb\\
F^{(1)}_{12}(x) &\equiv& \frac{1}{\pi}\left[1 + \frac{1}{4} x
+ \frac{9}{64} x^{2} +  O\left(x^{3}\right)\right],\nb\\
C_{0} &\equiv & \sum^{\infty}_{n = 0}{\frac{1}{(n + 1)(n+2)}}.
\eqn
Then, it can be shown that
\bq
\lb{4.29}
V_{1}(z) \rightarrow \frac{2c}{\pi}\ln(x) + O\left(x\ln(x)\right),
\;\;\; (k = 1),
\eq
as $x \rightarrow 0$. Thus, the analytical condition of $V_{1}(z)$ across the
hypersurface $x = 0 \; (z =1)$ requires $c = 0$. It can be further shown that
\bq
\lb{4.29a}
\varphi_{1}(z) \rightarrow a^{(2)}_{2}\ln(x) + O\left(x\ln(x)\right),
\;\;\; (k = 1),
\eq
as $x \rightarrow 0$. Thus, the analytical condition of $\varphi_{1}(z)$ across the
hypersurface $x = 0 $ requires $a^{(2)}_{2} = 0$. Considering Eqs.(\ref{4.13})
and (\ref{4.15}) we find that the boundary conditions in the present case limit all
the perturbations, that is,
\bq
\lb{4.30}
M_{1}(z) = S_{1}(z) =  V_{1}(z) = \varphi_{1}(z) = 0, \;\;\;
(k = 1).
\eq

Therefore, from Eq.(\ref{4.26}) we can see that for any given $n$, the solution has
\bq
\lb{4.31}
N = n -3, \;\;\;\;  (0 < 2q^{2} < 1),
\eq
unstable modes. In particular, the solution with
$n = 2$ or $n = 3$ has no unstable mode, and consequently  is stable with respect to
the linear perturbations. The solution with $n = 4$ has only one unstable mode, which
may represent a critical solution, sitting on a boundary that separates two attractive
basins in the phase space. All the solutions with $n > 4$
have more than one unstable modes and are not
stable with respect to the linear perturbations.

\subsection{$ 2 q^{2} \ge 1$}

In this case the boundary conditions at $z = 1$ are those given by Eq.(\ref{cd7}).
From Eq.(\ref{4.4}) we can see that to have $S_{1}(z)$ be finite as $z \rightarrow 1$,
we must assume that $Re(z) < 2$. To study the boundary conditions further,
let us first   consider the case $k \not=1$.

When $ k \not= 1$, it can be shown that  Eqs.(\ref{4.19})-(\ref{4.22c}) also hold
in the present case. Thus, the conditions that $V_{1}(z)$ and $\varphi_{1}(z)$ are finite
as $z \rightarrow 1$ require that
\bq
\lb{4.32}
Re(k) < 1,
\eq
while the condition that $M_{1}(z)$ is finite further requires that the constant
$a^{(2)}_{2}$  has to take the values given by Eq.(\ref{4.24}). On the other hand,
from Eqs.(\ref{4.22a})-(\ref{4.22c}) we can see that the last two conditions of
Eq.(\ref{cd7})  require
\bq
\lb{4.33}
Re(k) < 2q^{2} - 1.
\eq

When $ k = 1$, it can be shown that  Eqs.(\ref{4.27})-(\ref{4.29a})  hold in this
case, too, and following the analysis given there we find that the boundary conditions
at $z = 1$ limit all the perturbations for $k = 1$.

Therefore, for any given $q$ with $2q^{2} > 1$, Eqs.(\ref{4.32}) and (\ref{4.33})
show that there always exists a continuous spectrum of $k$ such that
\bq
\lb{4.34}
0 < Re(k) < Re(k_{min.}), \;\;\; (2q^{2} > 1),
\eq
where
\bq
\lb{4.35}
 Re(k_{min.}) = \cases{1, &  $q^{2} > 1$,\cr
  2q^{2} - 1, & $\frac{1}{2} < q^{2} < 1$.\cr}
\eq
That is, in the case $2q^{2} > 1$, there are infinite numbers of unstable modes.
However, when $2q^{2} = 1$,
from Eq.(\ref{4.33}) we find that
\bq
\lb{4.36}
Re(k) <  0, \;\;\; (2q^{2} = 1).
\eq
Thus, the solution with $2q^{2} = 1$ is stable against the linear perturbations.

\section{Summary and Concluding Remarks}

In this paper we have first introduced the notion of homothetic self-similarity to
four-dimensional spacetimes with cylindrical symmetry, and then presented a class of
exact solutions to the Einstein-massless scalar field equations, which is parameterized
by a constant, $q$. It has been shown that for $0 < 2q^{2} < 1$, the corresponding
spacetimes have black hole structures but with cylindrical symmetry. These black holes
are formed from the gravitational collapse of a massless scalar field. When $2q^{2} \ge 1$
the corresponding solutions also represent gravitational collapse of the scalar field
but no black holes are formed. Instead, a point-like singularity is developed, which is not naked
and can be seen by an observer only when he$/$she arrives at the singularity.

Then, the linear perturbations of all these solutions have been given analytically in closed
form in terms of hypergeometric functions. After properly imposing boundary conditions at
the axis and on the horizons, it has been shown that the solutions with $n = 2, 3$ and
the one with $2q^{2} = 1$ are stable, where $n$ is an integer and given by
$n \equiv 1/(1 - 2q^{2})$. For any given $n \ge 4$, the corresponding solution has $N = n -3$
unstable modes. In particular, the one with $n = 4$ has precisely one unstable mode, which
may represent a critical solution sitting on a boundary that separates two attractive
basins in the phase space. The solution for any given $q$ with $2q^{2} > 0$ has a continuous
spectrum of unstable eigen-modes, given by Eqs.(\ref{4.34}) and (\ref{4.35}).

It should be noted that in this paper we have shown that black holes can be formed from
gravitational collapse of the massless scalar field, and the ones with $n = 2, 3$
are stable against the linear perturbations. However, these spacetimes are not
asymptotically flat in the radial direction, and thus may not be considered as
representing counter-examples to the hoop conjecture \cite{Thorne72}. To have an
asymptotically flat spacetime in the radial direction, we may restrict the distribution
of the  scalar field only to a finite region, say, $r \le r_{0}(t)$, and then join it
with an asymptotically flat region.

\section*{Acknowledgments}

The author  would like to thank  M. Choptuik and N. Goldenfeld for their
carefully reading the manuscript and valuable discussions and comments. He would
also like to thank E. Hirschmann for delighted correspondence in critical collapse.
The Department of Physics, UIUC, is also greatly acknowledged for the hospitality.
The work is partially supported by CAPES, Brazil.

%%%%%%%%%%%%%%%%%%%%%%%%%%
\section*{Appendix A: The Ricci Tensor and Its Linear Perturbations in Terms of
Self-Similar Variables}

\lb{SecA}
\renewcommand{\theequation}{A.\arabic{equation}}
\setcounter{equation}{0}

The general metric for cylindrical spacetimes with two hypersurface orthogonal Killing
vectors takes the form of Eq.(\ref{2.1}).
%
%, which is
%%%%%%%%%%%%%%%%%%%%%%%%%%%%%%%%%%%%%%%%%%%%%%%%%%%%%%%%%%%%%%%%%%%%%%%%%%%%%%%%%%%%%%
%\bq
%\lb{A.1}
%ds^{2} = e^{-M(t,r)}\left(dt^{2} - dr^{2}\right)
%- r^{2}e^{-S(t,r)}\left(e^{V(t,r)}dw^{2}
%+ e^{-V(t,r)}d\theta^{2}\right).
%\eq
%where $x^{\mu} = \{t, r, \xi, \theta\}$, and the hypersurfaces $\theta = 0, 2\pi$
%are identified. Clearly, the two Killing vectors are given by
%$\xi_{(2)} = \partial_{\xi}$ and $\xi_{(3)} = \partial_{\theta}$.
%%%%%%%%%%%%%%%%%%%%%%%%%%%%%%%%%%%%%%%%%%%%%%%%%%%%%%%%%%%%%%%%%%%%%%%%%%%%%%%%%%%%%%
%
Introducing the self-similar dimensionless variables $\tau$ and $z$ vis the relations,
\bq
\lb{A.2}
\tau = - \ln\left(-\frac{t}{t_{0}}\right),\;\;\;\;
z = \frac{r}{(-t)},
\eq
where $t_{0}$ is a dimensional constant, we find that the non-vanishing components of
the Ricci tensor are given by,
\bqn
\lb{A.3}
%\lb{A.3a}
R_{tt} &=& \frac{e^{2\tau}}{2 z t_{0}^{2}}\left\{2z^{3}S_{,zz}
- \left(1 - z^{2}\right)\left(zM_{,zz} + 2M_{,z}\right) - z^{3}\left({S_{,z}}^{2}
+ {V_{,z}}^{2}\right) + z \left(1 + z^{2}\right)M_{,z}S_{,z}
 + 4 z^{2}S_{,z}\right.\nb\\
 & & + z\left[2S_{,\tau\tau} + M_{,\tau\tau}
 - \left({S_{,\tau}}^{2}  + {V_{,\tau}}^{2}\right)
 + \left(2 + M_{\tau}\right)S_{,\tau} + M_{,\tau}\right]\nb\\
 & & \left. + z^{2}\left[4S_{,\tau z} + 2M_{,\tau z} - 2\left(S_{,\tau} S_{,z}
 + V_{,\tau} V_{,z}\right) + \left(M_{,\tau} S_{,z}
 + M_{,z} S_{,\tau}\right)\right]\right\},\nb\\
%
%
%\lb{A.3b}
R_{tr} &=& \frac{e^{2\tau}}{2 z t_{0}^{2}}\left\{2z^{2}S_{,zz}
 - z^{2}\left({S_{,z}}^{2} + {V_{,z}}^{2}\right) - 2z\left(1 - zS_{,z}\right)M_{,z}
 + 4z S_{,z}\right.\nb\\
 & & \left. + z\left[2 S_{,\tau z} - \left(S_{,\tau} S_{,z}
 + V_{,\tau} V_{,z}\right) + \left(M_{,\tau} S_{,z}
 + M_{,z} S_{,\tau}\right)\right]  + 2\left(S_{,\tau}- M_{,\tau}\right)\right\},\nb\\
%
%
%\lb{A.3c}
R_{rr} &=& \frac{e^{2\tau}}{2 z t_{0}^{2}}\left\{2zS_{,zz}
  + z\left(1 - z^{2}\right)M_{,zz} - z\left({S_{,z}}^{2} + {V_{,z}}^{2}\right)
+ z \left(1 + z^{2}\right)M_{,z}S_{,z} - 2\left(1 + z^{2}\right)M_{,z}
 + 4 S_{,z}\right.\nb\\
 & & \left. - z\left[M_{,\tau\tau}+ \left(1 - {S_{,\tau}}\right)M_{,\tau}\right]
 - z^{2}\left[2M_{,\tau z} -  \left(M_{,\tau} S_{,z}
 + M_{,z} S_{,\tau}\right)\right]\right\},\nb\\
%
%
%\lb{A.3d}
R_{22} &=& \frac{1}{2}e^{M + V - S}\left\{z^{2}\left(1 - z^{2}\right)
\left[\left(S_{,zz}
- V_{,zz}\right) - S_{,z}\left(S_{,z} - V_{,z}\right)\right]
+ 2z \left(1 - z^{2}\right)\left(S_{,z} - V_{,z}\right)
-  2\left(1 - zS_{,z}\right)\right.\nb\\
& & - z^{3}\left[2\left(S_{,\tau z} - V_{,\tau z}\right) - 2S_{,\tau}S_{,z}
+ \left(S_{,\tau} V_{,z}  + S_{,z} V_{,\tau}\right)\right]\nb\\
& &\left.  - z^{2}\left[\left(S_{,\tau\tau} - V_{,\tau\tau}\right)
+ \left(1 - S_{,\tau}\right)\left(S_{,\tau}-V_{,\tau}\right)\right]\right\},\nb\\
%
%
%\lb{A.3e}
R_{33} &=& \frac{1}{2}e^{M - V - S}\left\{z^{2}\left(1 - z^{2}\right)
\left[\left(S_{,zz}
  + V_{,zz}\right) - S_{,z}\left(S_{,z} + V_{,z}\right)\right]
+ 2z \left(1 - z^{2}\right)\left(S_{,z} + V_{,z}\right)
-  2\left(1 - zS_{,z}\right)\right.\nb\\
& & - z^{3}\left[2\left(S_{,\tau z} + V_{,\tau z}\right) - 2S_{,\tau}S_{,z}
- \left(S_{,\tau} V_{,z}  + S_{,z} V_{,\tau}\right)\right]\nb\\
& &\left.  - z^{2}\left[\left(S_{,\tau\tau} + V_{,\tau\tau}\right)
+ \left(1 - S_{,\tau}\right)\left(S_{,\tau} + V_{,\tau}\right)\right]\right\}.
\eqn

%where $(\;)_{,a} \equiv \partial(\;)/\partial x^{a}$.

On the other hand, it can be shown that the Klein-Gordon equation, $\Box \phi = 0$,
for the massless scalar field takes the form
\bqn
\lb{KG}
z(1-z^{2})\phi_{,zz} - 2z^{2}\phi_{,\tau z} - z \phi_{,\tau\tau}
&+& \left[z(zS_{,z} - 1) + zS_{,\tau}\right]  \phi_{,\tau}\nb\\
&+& \left[(1-z^{2})(2- zS_{,z}) + z^{2}S_{,\tau}\right]  \phi_{,z} = 0.
\eqn

Now let us consider the linear perturbations of Eq.(\ref{4.1}).
To first order in $\epsilon$, it can be shown that the non-vanishing
components of the Ricci tensor are given by
\bqn
\lb{A.5a}
R_{tt}^{(1)} &=& \frac{e^{(2 + k)\tau}}{2 z t_{0}^{2}}\left\{2z^{3}S''_{1}
- z\left(1 - z^{2}\right)M''_{1} - 2z^{3}V'_{0}V'_{1}
- \left[2 - 2(1+k)z^{2} - z\left(1 + z^{2}\right)S'_{0}\right]M'_{1}\right.\nb\\
& &  + z\left[4(1+k)z + \left(1 + z^{2}\right)M'_{0} - 2z^{2}S'_{0}\right]S'_{1}
- 2kz^{2}V'_{0}V_{1} + kz\left(1+k + zS'_{0}\right)M_{1}\nb\\
& & \left.
+ kz \left[2(1+k) + z\left(M'_{0} - 2S'_{0}\right)\right]S_{1}\right\},\\
\lb{A.5b}
R_{tr}^{(1)} &=& \frac{e^{(2 + k)\tau}}{2 z t_{0}^{2}}\left\{2z^{2}S''_{1}
- 2z^{2}V'_{0}V'_{1} + 2z\left[(2+k) + z\left(M'_{0} - S'_{0}\right)\right]S'_{1}
\right.\nb\\
& & \left. - 2z\left(1 - zS'_{0}\right)M'_{1}
- kzV'_{0}V_{1} - k\left(2- zS'_{0}\right)M_{1}
+ k\left[2  + z\left(M'_{0} - S'_{0}\right)\right]S_{1}\right\},\\
\lb{A.5c}
R_{rr}^{(1)} &=& \frac{e^{(2 + k)\tau}}{2 z t_{0}^{2}}\left\{2zS''_{1}
+ z\left(1 - z^{2}\right)M''_{1} - 2zV'_{0}V'_{1}
+\left[4 +  z\left(1 + z^{2}\right)M'_{0} - 2zS'_{0}\right]S'_{1}\right.\nb\\
& & \left. - \left[\left(1 + z^{2}\right)\left(2 - zS'_{0}\right)
+ 2k z^{2}\right]M'_{1}
-  kz \left(1 + k - zS'_{0}\right)M_{1}
+ k z^{2}M'_{0}S_{1}\right\},\\
\lb{A.5d}
R_{22}^{(1)} &=&  e^{k\tau}\left\{\left(M_{1} + V_{1} - S_{1}\right) R^{(0)}_{22}
+ \frac{1}{2}ze^{M_{0}  + V_{0} - S_{0}}\bar{R}^{(1)}_{22}\right\},\\
\lb{A.5e}
R_{33}^{(1)} &=&  e^{k\tau}\left\{\left(M_{1} - V_{1} - S_{1}\right) R^{(0)}_{33}
+ \frac{1}{2}ze^{M_{0}  - V_{0} - S_{0}}\bar{R}^{(1)}_{33}\right\},
\eqn
where $R^{(0)}_{22}$ and $R^{(0)}_{33}$ are the corresponding components  of the
Ricci tensor for the background solution, and
\bqn
\lb{A.6a}
\bar{R}^{(1)}_{22} &=& z\left(1-z^{2}\right)\left(S''_{1} - V''_{1}\right)
+ \left[2kz^{2} + \left(1 - z^{2}\right)\left(zS'_{0} - 2\right)\right]V'_{1}
+ \left[4 - 2(1+k)z^{2} - z\left(1-z^{2}\right)
\left(2S'_{0} - V'_{0}\right)\right]S'_{1}\nb\\
& & + kz\left(1 + k - zS'_{0}\right)V_{1}
- kz\left[(1+k) - z\left(2S'_{0} - V'_{0}\right)\right]S_{1},\\
\lb{A.6b}
\bar{R}^{(1)}_{33} &=& z\left(1-z^{2}\right)\left(S''_{1} + V''_{1}\right)
- \left[2kz^{2} + \left(1 - z^{2}\right)\left(zS'_{0} - 2\right)\right]V'_{1}
+ \left[4 - 2(1+k)z^{2} - z\left(1-z^{2}\right)
\left(2S'_{0} + V'_{0}\right)\right]S'_{1}\nb\\
& & - kz\left(1 + k - zS'_{0}\right)V_{1}
- kz\left[(1+k) - z\left(2S'_{0} + V'_{0}\right)\right]S_{1}.
\eqn

\section*{Appendix B: Apparent Horizons in Spacetimes with Cylindrical Symmetry}

\lb{SecB}
\renewcommand{\theequation}{B.\arabic{equation}}
\setcounter{equation}{0}

In \cite{LW94}, the ingoing and outgoing radial null geodesics were studied in
double null coordinates, and the corresponding  expansions of them
were calculated.  In this paper,  we shall define apparent horizons in spacetimes with
cylindrical symmetry, using those quantities. Before doing so, we would like
first to note the difference  between the double null coordinates used in this
paper and the ones used in \cite{LW94}. As a matter of fact, the roles of $u$
and $v$ are exchanged in this paper.

To write the metric (\ref{2.1}) in the double null coordinates, let us first
introduce the two null coordinates $u$ and $v$ via the relations,
\bq
\lb{B.1}
t = \alpha(u) + \beta(v),\;\;\;
r = \alpha(u) - \beta(v),
\eq
where $\alpha(u)$ and  $\beta(v)$ are two arbitrary functions of their indicated
arguments, subject to
\bq
\lb{B.1a}
 \alpha'(u)\beta'(v) \not= 0,
 \eq
where a prime denotes the ordinary differentiation. Then, in terms of $u$ and $v$,
the metric (\ref{2.1}) takes the form
  \bq
  \lb{B.2}
  ds^{2} = 2 e^{2\sigma(u, v)} dudv - r^{2}e^{-S(t,r)}\left(e^{V(t,r)}dw^{2}
+ e^{-V(t,r)}d\theta^{2}\right),
  \eq
 where
  \bq
  \lb{B.3}
  \sigma(u,v) =  \frac{1}{2}\left[\ln\left(2\alpha'\beta'\right)
  - M\right].
  \eq

Introducing two null vectors $l^{\lambda}$ and $n^{\lambda}$  by
 \bq
 \lb{B.4}
 l_{\lambda} \equiv \frac{\partial u}{\partial x^{\lambda}} =
 \delta^{u}_{\lambda},\;\;\;
 n_{\lambda} \equiv \frac{\partial v}{\partial x^{\lambda}} =
 \delta^{v}_{\lambda},
 \eq
 we find that
 \bq
 \lb{B.5}
 l_{\mu;\nu} l^{\nu} = 0 = n_{\mu;\nu} n^{\nu},
 \eq
 which means that each of them defines an affinely parametrized null
 geodesic congruence. In particular,  $l^{\mu}$ defines the one moving along the null
 hypersurfaces $u =  Const.$, while $n^{\mu}$ defines the one
 moving along the null hypersurfaces $v =  Const.$
  Then, the expansions of these null geodesics are defined as,
  \bqn
  \lb{B.6}
\Theta_{l} &\equiv& g^{\alpha\beta} l_{\alpha;\beta}
   = e^{-2\sigma}\frac{{\cal{R}}_{,v}}{{\cal{R}}},\nb\\
\Theta_{n} &\equiv& g^{\alpha\beta} n_{\alpha;\beta}
  = e^{-2\sigma}\frac{{\cal{R}}_{,u}}{{\cal{R}}},
  \eqn
where
\bq
\lb{B.7}
{\cal{R}} \equiv \left||\partial_{w} \cdot \partial_{w}\right|| \cdot
\left||\partial_{\theta}
\cdot \partial_{\theta}\right||   = r^{2} e^{-S}.
\eq

We call the cylinders of constant $t$ and $r$ trapped if $\Theta_{l}\Theta_{n} > 0$,
marginally trapped if $\Theta_{l}\Theta_{n} = 0$, and untrapped if $\Theta_{l}\Theta_{n}
< 0$. An apparent horizon is defined as a hypersurface foliated by marginally trapped
surfaces \cite{HE73,Hay94}.

\end{document}